\documentstyle[preprint,eqsecnum,epsf,aps,floats,tighten]{revtex}
\begin{document}
\renewcommand{\baselinestretch}{1.0}
% Change equation numbers to single arabic
\def\theequation{\arabic{equation}}%
\newcommand{\rstev}{\mbox{$\rs = \T{1.8}$}}
\newcommand{\AP}{\mbox{${\rm \bar{p}}$}}
\newcommand{\SU}{\mbox{$S$}}
\newcommand{\SPt}{\mbox{$<\! |S|^2 \!>$}}
\newcommand{\ET}{\mbox{$E_{T}$}}
\newcommand{\PT}{\mbox{$p_{t}$}}
\newcommand{\DP}{\mbox{$\Delta\phi$}}
\newcommand{\DR}{\mbox{$\Delta R$}}
\newcommand{\DE}{\mbox{$\Delta\eta$}}
\newcommand{\DEP}{\mbox{$\Delta\eta_{c}$}}
\newcommand{\DEC}{\mbox{$\Delta\eta_{c}$}}
\newcommand{\SP}{\mbox{$S(\DEP)$}}
\newcommand{\PH}{\mbox{$\phi$}}
\newcommand{\EA}{\mbox{$\eta$}}
\newcommand{\EAJ}{\mbox{\EA(jet)}}
\newcommand{\AEA}{\mbox{$|\eta|$}}
\newcommand{\Ge}[1]{\mbox{#1 GeV}}
\newcommand{\T}[1]{\mbox{#1 TeV}}
\newcommand{\D}[1]{\mbox{$#1^{\circ}$}}
\newcommand{\x}{\cdot}
\newcommand{\ra}{\rightarrow}
% units
\newcommand{\mb}{\mbox{mb}}
\newcommand{\nb}{\mbox{nb}}
\newcommand{\ipb}{\mbox{${\rm pb}^{-1}$}}
\newcommand{\inb}{\mbox{${\rm nb}^{-1}$}}
\newcommand{\rs}{\mbox{$\sqrt{s}$}}
\newcommand{\fdel}{\mbox{$f(\DEP)$}}
\newcommand{\fdele}{\mbox{$f(\DEP)^{exp}$}}
\newcommand{\fgap}{\mbox{$f(\DEP\! > \!3)$}}
\newcommand{\fgape}{\mbox{$f(\DEP\! > \!3)^{exp}$}}
\newcommand{\fpyt}{\mbox{$f(\DEP\!>\!2)$}}
\newcommand{\delth}{\mbox{$\DEP\! > \!3$}}
\newcommand{\uplim}{\mbox{$1.1\!\times\!10^{-2}$}}
\newcommand{\sigew}{\mbox{$\sigma_{\rm EW}$}}
\newcommand{\sigdi}{\mbox{$\sigma_{\rm dijet}$}}
\newcommand{\sigsi}{\mbox{$\sigma_{\rm singlet}$}}
\newcommand{\sigr}{\mbox{$\sigsi/\sigma$}}
\newcommand{\sigrew}{\mbox{$\sigew/\sigma$}}
\newcommand{\ncal}{\mbox{$n_{\rm cal}$}}
\newcommand{\ntrk}{\mbox{$n_{\rm trk}$}}
% greater/less than or approximately equal to 
\def\simge
{\mathrel{\rlap{\raise 0.53ex \hbox{$>$}}{\lower 0.53ex \hbox{$\sim$}}}}
\def\simle
{\mathrel{\rlap{\raise 0.4ex \hbox{$<$}}{\lower 0.72ex \hbox{$\sim$}}}}
%
% D0 Style Definitions
% Definitions of commonly used symbols
%
\def\sigtot{$\sigma_{\rm tot}$}         %sigma total
\def\sigtop{$\sigma_{t \overline{t}}$}  %sigma_ttbar
\def\pbarp{$\overline{p}p $}            %pbarp
\def\ppbar{$p\overline{p} $}            %ppbar
\def\qqbar{$q\overline{q}$}             %qqbar
\def\qbarq{$\overline{q}q$}             %qbarq
\def\ttbar{$t\overline{t}$}             %ttbar
\def\bbbar{$b\overline{b}$}             %bbbar
\def\D0{D\O}                            %D0
\def\CDF{CDF}
\def\ipb{pb$^{-1}$}                     %inverse picobarns
\def\pt{p_T}                            %pT
\def\ptg{p_T^\gamma}                    %pT_gamma
\def\et{E_T}                            %ET
\def\etg{E_T^\gamma}                    %ET_gamma
\def\htran{$H_T$}                       %HT
\def\gevcc{{\rm GeV}/c$^2$}                   %GeV/c^2
\def\gevc{{\rm GeV}/c}                  %GeV/c
\def\gev{~\rm GeV}                       %GeV
\def\tev{~\rm TeV}                       %TeV
\def\njet{$N_{\rm jet}$}                %N_jet
\def\aplan{$\cal{A}$}                   %aplanarity
\def\lum{$\cal{L}$}                     %luminosity
\def\iso{$\cal{I}$}                     %isolation variable
\def\remu{${\cal{R}}_{e\mu}$}           %distance between e and mu in eta-phi
\def\rmu{$\Delta\cal{R}_{\mu}$}         %distance between mu and jet cone axis
\def\pbar{$\overline{p}$}               %pbar
\def\tbar{$\overline{t}$}               %tbar
\def\bbar{$\overline{b}$}               %bbar
\def\lumint{$\int {\cal{L}} dt$}        %integrated luminosity
\def\lumunits{cm$^{-2}$s$^{-1}$}        %luminosity units
\def\etal{{\sl et al.}}                 %et al. - no preceeding comma
\def\vs{{\sl vs.}}                      %vs. 
\def\sinthw{sin$^2 \theta_W$}           %sin^2 th_W
\def\mt{$m_t$}                          %m_top 
\def\mb{$m_b$}                          %m_bottom
\def\mw{$M_W$}                          %M_W
\def\mz{$M_Z$}                          %M_Z
\def\pizero{$\pi^0$}                    %pizero
\def\jpsi{$J/\psi$}                     %J/psi
\def\wino{$\widetilde W$}               %Wino
\def\zino{$\widetilde Z$}               %Zino
\def\squark{$\widetilde q$}             %squark
\def\gluino{$\widetilde g$}             %gluino
\def\alphas{$\alpha_{\scriptscriptstyle S}$}                %alpha_s
\def\alphaem{$\alpha_{\scriptscriptstyle{\rm EM}}$}         %alpha_EM
\def\epm{$e^+e^-$}                      %e+e-
\def\deg{$^\circ$}                      %degree sign
\def\met{\mbox{${\hbox{$E$\kern-0.6em\lower-.1ex\hbox{/}}}_T$ }} %missing ET
\def\mht{\mbox{${\hbox{$H$\kern-0.725em\lower-.1ex\hbox{/}}}_T$}} %missing HT
\def\st{\mbox{$S_T$}}          % same as \mht
\def\gup{{g\uparrow}} 
\def\gdown{{g\downarrow}} 
%%
%%
%% \def\draft{\special{header=draft.ps}}
%%      
%%  References help:  DO NOT USE THESE FOR PAPERS WHICH WILL BE SUBMITTED
%%                               TO PRL OR PRD!!!!!!!!!!
%%
\newcommand{\NC}{{\em Nuovo Cimento\/} }
\newcommand{\NIM}{{\em Nucl. Instr. Meth.} }
\newcommand{\NP}{{\em Nucl. Phys.} }
\newcommand{\PL}{{\em Phys. Lett.} }
\newcommand{\PR}{{\em Phys. Rev.} }
\newcommand{\PRL}{{\em Phys. Rev. Lett.} }
\newcommand{\RMP}{{\em Rev. Mod. Phys.} }
\newcommand{\ZP}{{\em Zeit. Phys.} }
% Parameters for references:
% parameter 1 - volume, parameter 2 - page, parameter 3- last two digits of year
\def\err#1#2#3 {{\it Erratum} {\bf#1},{\ #2} (19#3)}
\def\ib#1#2#3 {{\it ibid.} {\bf#1},{\ #2} (19#3)}
\def\nc#1#2#3 {Nuovo Cim. {\bf#1} ,#2(19#3)}
\def\nim#1#2#3 {Nucl. Instr. Meth. {\bf#1},{\ #2} (19#3)}
\def\np#1#2#3 {Nucl. Phys. {\bf#1},{\ #2} (19#3)}
\def\pl#1#2#3 {Phys. Lett. {\bf#1},{\ #2} (19#3)}
\def\prev#1#2#3 {Phys. Rev. {\bf#1},{\ #2} (19#3)}
\def\prl#1#2#3 {Phys. Rev. Lett. {\bf#1},{\ #2} (19#3)}
\def\rmp#1#2#3 {Rev. Mod. Phys. {\bf#1},{\ #2} (19#3)}
\def\zp#1#2#3 {Zeit. Phys. {\bf#1},{\ #2} (19#3)}

%\tightenlines
\hyphenation{PYTHIA}
\lefthyphenmin=3
\righthyphenmin=3
%\preprint{Fermilab-Pub-99/357-E}
%\preprint{ }

\title{Limits on Quark Compositeness from High Energy Jets 
                        in $\bbox{\bar{p}p}$  Collisions at 1.8 TeV}

%% file: list_of_authors.tex                  
% LIST_OF_AUTHORS.TEX                 4/12/00            
%
\author{                                                                      
%% names begin here                                                           
B.~Abbott,$^{48}$                                                             
M.~Abolins,$^{45}$                                                            
V.~Abramov,$^{21}$                                                            
B.S.~Acharya,$^{15}$                                                          
D.L.~Adams,$^{55}$                                                            
M.~Adams,$^{32}$                                                              
V.~Akimov,$^{19}$                                                             
G.A.~Alves,$^{2}$                                                             
N.~Amos,$^{44}$                                                               
E.W.~Anderson,$^{37}$
% ...
\\                                                                            
\vskip 0.30cm                                                                 
\centerline{(D\O\ Collaboration)}                                             
\vskip 0.30cm                                                                 
}                                                                             
\address{                                                                     
\centerline{$^{1}$Universidad de Buenos Aires, Buenos Aires, Argentina}       
\centerline{$^{2}$LAFEX, Centro Brasileiro de Pesquisas F{\'\i}sicas,         
                  Rio de Janeiro, Brazil}                                     
\centerline{$^{3}$Universidade do Estado do Rio de Janeiro,                   
                  Rio de Janeiro, Brazil}                                     
\centerline{$^{4}$Institute of High Energy Physics, Beijing,                  
                  People's Republic of China}                                 
\centerline{$^{5}$Universidad de los Andes, Bogot\'{a}, Colombia}             
\centerline{$^{6}$Charles University, Prague, Czech Republic}                 
\centerline{$^{7}$Institute of Physics, Academy of Sciences, Prague,          
                  Czech Republic}                                             
\centerline{$^{8}$Universidad San Francisco de Quito, Quito, Ecuador}         
\centerline{$^{9}$Institut des Sciences Nucl\'eaires, IN2P3-CNRS,             
                  Universite de Grenoble 1, Grenoble, France}                 
\centerline{$^{10}$CPPM, IN2P3-CNRS, Universit\'e de la M\'editerran\'ee,     
                  Marseille, France}                                          
\centerline{$^{11}$LPNHE, Universit\'es Paris VI and VII, IN2P3-CNRS,         
                  Paris, France}                                              
\centerline{$^{12}$DAPNIA/Service de Physique des Particules, CEA, Saclay,    
                  France}                                                     
\centerline{$^{13}$Panjab University, Chandigarh, India}                      
\centerline{$^{14}$Delhi University, Delhi, India}                            
\centerline{$^{15}$Tata Institute of Fundamental Research, Mumbai, India}     
\centerline{$^{16}$Seoul National University, Seoul, Korea}                   
\centerline{$^{17}$CINVESTAV, Mexico City, Mexico}                            
\centerline{$^{18}$Institute of Nuclear Physics, Krak\'ow, Poland}            
\centerline{$^{19}$Institute for Theoretical and Experimental Physics,        
                   Moscow, Russia}                                            
\centerline{$^{20}$Moscow State University, Moscow, Russia}                   
\centerline{$^{21}$Institute for High Energy Physics, Protvino, Russia}       
\centerline{$^{22}$Lancaster University, Lancaster, United Kingdom}           
\centerline{$^{23}$University of Arizona, Tucson, Arizona 85721}              
\centerline{$^{24}$Lawrence Berkeley National Laboratory and University of    
                  California, Berkeley, California 94720}                     
\centerline{$^{25}$University of California, Davis, California 95616}         
\centerline{$^{26}$California State University, Fresno, California 93740}     
\centerline{$^{27}$University of California, Irvine, California 92697}        
\centerline{$^{28}$University of California, Riverside, California 92521}     
\centerline{$^{29}$Florida State University, Tallahassee, Florida 32306}      
\centerline{$^{30}$University of Hawaii, Honolulu, Hawaii 96822}              
\centerline{$^{31}$Fermi National Accelerator Laboratory, Batavia,            
                   Illinois 60510}                                            
\centerline{$^{32}$University of Illinois at Chicago, Chicago,                
                   Illinois 60607}                                            
\centerline{$^{33}$Northern Illinois University, DeKalb, Illinois 60115}      
\centerline{$^{34}$Northwestern University, Evanston, Illinois 60208}         
\centerline{$^{35}$Indiana University, Bloomington, Indiana 47405}            
\centerline{$^{36}$University of Notre Dame, Notre Dame, Indiana 46556}       
\centerline{$^{37}$Iowa State University, Ames, Iowa 50011}                   
\centerline{$^{38}$University of Kansas, Lawrence, Kansas 66045}              
\centerline{$^{39}$Kansas State University, Manhattan, Kansas 66506}          
\centerline{$^{40}$Louisiana Tech University, Ruston, Louisiana 71272}        
\centerline{$^{41}$University of Maryland, College Park, Maryland 20742}      
\centerline{$^{42}$Boston University, Boston, Massachusetts 02215}            
\centerline{$^{43}$Northeastern University, Boston, Massachusetts 02115}      
\centerline{$^{44}$University of Michigan, Ann Arbor, Michigan 48109}         
\centerline{$^{45}$Michigan State University, East Lansing, Michigan 48824}   
\centerline{$^{46}$University of Nebraska, Lincoln, Nebraska 68588}           
\centerline{$^{47}$Columbia University, New York, New York 10027}             
\centerline{$^{48}$New York University, New York, New York 10003}             
\centerline{$^{49}$University of Rochester, Rochester, New York 14627}        
\centerline{$^{50}$State University of New York, Stony Brook,                 
                   New York 11794}                                            
\centerline{$^{51}$Brookhaven National Laboratory, Upton, New York 11973}     
\centerline{$^{52}$Langston University, Langston, Oklahoma 73050}             
\centerline{$^{53}$University of Oklahoma, Norman, Oklahoma 73019}            
\centerline{$^{54}$Brown University, Providence, Rhode Island 02912}          
\centerline{$^{55}$University of Texas, Arlington, Texas 76019}               
\centerline{$^{56}$Texas A\&M University, College Station, Texas 77843}       
\centerline{$^{57}$Rice University, Houston, Texas 77005}                     
\centerline{$^{58}$University of Washington, Seattle, Washington 98195}       
}                                                                             
%end                                                                          

\maketitle
\pagebreak

\renewcommand{\baselinestretch}{1.5}
\normalsize
\begin{abstract}
Events in \pbarp\ collisions at \rstev\ with total transverse energy
exceeding $500 \gev$ are used to set limits on quark substructure.  
The data are consistent with next-to-leading order QCD calculations.
We set a lower limit of $2.0 \tev$ at 
95\% confidence on the energy scale 
$\Lambda_{\it LL}$ for compositeness in quarks, 
assuming a model with a
left-left isoscalar contact interaction term.   The limits on 
$\Lambda_{\it LL}$ are found to be insensitive to the sign of the 
interference term in the Lagrangian.
\end{abstract}

%\twocolumn
%PAGE 5, PARA 1
The first limit on the size of the atomic nucleus was obtained by Geiger
and Marsden in the Rutherford \cite{Rutherford} scattering
of $\alpha$ particles from nuclei.  In an analogous way, we can set a
limit on the size of quarks by observing the scattering of the highest 
energy quarks and antiquarks at the Fermilab Tevatron Collider at
\pbarp\ center-of-mass energies of $1.8 \tev$.  
The scattered quarks from within the proton emerge in the 
laboratory as collimated showers of hadrons, called jets.
The scalar sum of the transverse energies of the jets
in any event provides a measure of the hardness (the impact
parameter) of collisions.
The summed transverse energy of the event is simply expressed
$$H_T \equiv \sum_{i=1}^{N} {E_T^{i}\, ,}$$
where $N$ is the number of jets in the event above some threshold,
and $E_T^{i}$ is the transverse energy of jet $i$, essentially
the momentum component of the jet in the plane transverse to the beams
\cite{Baden}.

%PAGE 5, PARA 2
\htran\ is a robust quantity in the multiple interaction environment
of the Tevatron, where often a hard scattering is accompanied by one or more
soft interactions that do not produce high $E_T$ jets.
Such overlapping events contribute 
only a small and easily corrected bias to \htran.  
For individual jets, the precise measurement of the
hard-scattering vertex is crucial for determining $E_T^i$, but
changes in $E_T^i$ induced by changing the position of the vertex are 
partially compensated in \htran.
Efficiencies and resolutions are measured as functions of $E_T^i$; these
are correlated weakly with \htran\ because of an effective averaging over 
final-state topologies.  By treating the event as a whole, 
this analysis complements 
the more traditional probes of QCD, such as measurements of the 
inclusive jet cross section\cite{d0_inc,cdf_inc}, the dijet 
mass spectrum\cite{Mjj}, and the dijet angular 
distribution\cite{d0_dijet,cdf_dijet}. 
A measurement of $d\sigma / d H_T$ has been published by 
the CDF collaboration\cite{cdf_ht}.

%PAGE 6, PARA 1
This analysis focuses on a test of quark compositeness within the 
formalism of Eichten {\it et al.}\cite{theory-comp} for events with 
$H_T > 500$ \gev.  In the Lagrangian of Ref. \cite{theory-comp}, 
we test for compositeness of left-handed quarks in the 
left-left isoscalar term,
$$L_{qq} = {\cal A}\, ( g^2 / 2 \Lambda_{\it LL}^2)\, \overline{q_{_L}} 
\gamma^\mu q_{_L} \overline{q_{_L}} \gamma_\mu q_{_L}\,,$$
where ${\cal A} = \pm 1$ is the sign of the interference term, 
$\Lambda_{\it LL}$ is the compositeness scale, and the 
dependence on $\alpha_s$ is contained in the compositeness 
coupling constant $g^2$.  The 
model is completely determined by specifying the two parameters $\cal{A}$ and
$\Lambda_{\it LL}$.  In this model, all three 
families of quarks are assumed to be composite, and 
both signs of the interference term (resulting in 
constructive ($-1$) and destructive ($+1$) interference) are investigated.  
In this search for quark compositeness at jet energies well above the 
mass of the top quark, with \mbox{$H_T > 500~ \rm GeV > 2 m_t \approx
350 \gev$}, the only backgrounds considered are from instrumental sources.
For comparison to these results, 
Table \ref{limits_to_date} shows the previous quark compositeness limits.

%PAGE 6, PARA 2
The \D0\ detector is described in detail in Ref. \cite{det}.
The principal components of the detector used in this analysis are the 
calorimeter for measuring jets, and the central tracking
system for determining the hard-scattering vertex.  
The pseudorapidity, $\eta = -\ln(\tan(\theta /2))$, of the calorimeter 
extends to $|\eta| \le$ 4.2, corresponding to a polar angle 
relative to the incident proton of $\theta \approx 2^{\circ}.$
The depth of the \D0\ calorimeter varies from
6 to 10 nuclear interaction lengths, thereby
providing good containment for jets.  Jet energy resolution is
approximately 80$\%/\sqrt{E}$, and the resolution on the 
$z$-position of the hard-scattering vertex is $\pm 8$ mm.

%PAGE 6, PARA 3
Our analysis is based on $91.9\pm5.6$ \ipb \cite{luminosity} of data taken 
during the 1994-1995 run of the Tevatron.  
The hardware trigger required a minimum transverse energy exceeding
$45 \gev$ in a region $\Delta \eta \times \Delta \phi = 0.8 \times 1.6$ of the
calorimeter, where $\phi$ is the azimuthal angle.  
In addition, beam halo effects from the Main Ring, the preaccelerator
to the Tevatron,
were minimized through
timing restrictions.  
The software filter required
at least one jet with $E_T > 115 \gev$. 
The combined selection efficiency was found to exceed 99\% for 
events with $H_T > 500 \gev$.

%PAGE 7, PARA 1
A significant fraction of the data were taken at high
instantaneous luminosity, which resulted in more than one
\pbarp\ interaction in a beam crossing leading to an ambiguity in
selecting the primary event vertex. After event
reconstruction, the two vertices with the largest track multiplicity
were retained. When there was a second reconstructed vertex in the event, 
the imbalance in transverse momentum or missing \ET ~(\met) 
was calculated using transverse vector 
energies: 
$$ \met \equiv \arrowvert \sum_{i=1}^{N} 
{\buildrel{\rightarrow}\over{E_T^i}} \arrowvert \, .$$
This was evaluated for both event vertex candidates,
with the primary vertex chosen to minimize \met.
The $z$-position of the vertex was required to satisfy 
$\vert z_{vtx} \vert \leq 50$ cm.  The efficiency for this 
cut was measured to be approximately $90$\%, independent of \htran.

%PAGE 7, PARA 2
Offline jet reconstruction used a fixed-cone 
algorithm with radius 
$${\cal{R}} = \sqrt{(\Delta \eta)^2 + (\Delta \phi)^2} = 0.7\,,$$ 
and was fully efficient for $\et > 20 \gev$, the threshold 
applied to each jet for inclusion in \htran.  
The jet energy scale corrections applied to the data 
are described in Ref. \cite{d0_escale}.
Additional offline cuts were applied to the events to 
minimize instrumental background 
and ambiguities in defining $E_T^i$ and \met.

%PAGE 8, PARA 1
All jets with $\et > 20 \gev$ and with $\vert\eta^j\vert < 3.0$
were required to pass 
jet selection criteria, which included: (i) the electromagnetic
fraction of the jet energy, measured in the first layers of the 
uranium-liquid-argon
calorimeter, was required to be between 0.05 and 0.95, except in the 
region between the central and end cryostats, where only the upper limit was 
imposed; (ii) the fraction of energy in the outermost hadronic 
section was required 
to be $<$ 0.40;
and, (iii) the ratio $E_T^{cell~1} / E_T^{cell~2}$ was required to be
$< 10$, where the calorimeter
cells comprising the jet were ordered in decreasing \ET.
An event was rejected if any of its jets with $E_T > 20 \gev$
failed the quality or $\eta$ requirements.  
The efficiency for a jet to pass these criteria 
was parameterized as a function of $E_T$, and the efficiency for an
event to pass the criteria was essentially independent 
of \htran\ above $500 \gev$.

%PAGE 8, PARA 2
The \htran\ distribution for $H_T > 500 \gev$ is shown in
Fig.\ref{dataxsection_figure}.  
The events passed all 
the above selection criteria and were corrected for efficiencies and
jet energy scale, but not for resolution.
The cross section falls by three orders-of-magnitude 
over the range in \htran\ from $500-1000 \gev$.  
Fig. \ref{dth_figure} displays 
the fractional deviation between the data and the Monte Carlo
for the CTEQ4M PDF with a 
renormalization scale of $E_T^{\rm max}/2$.

%PAGE 8, PARA 3
The \htran\ spectrum expected from the standard model was
provided by the {\sc jetrad}\cite{jetrad} 
Monte Carlo event generator, which is based on
a next-to-leading order (NLO) QCD calculation. 
We tried several choices for the
renormalization scale $\mu$ parameterized as $\mu = f_E \cdot E_T^{\rm max}$ 
and $\mu = f_H \cdot H_T$, where $f_E$ and $f_H$ are constants we 
varied from 0.25 to 1.50.  We used two parton distribution functions
(PDFs): CTEQ4M\cite{cteq} and MRST\cite{mrst}.

%PAGE 8, PARA 4
For $\Lambda_{\it LL}$ scales between 
$1.4$ and $7.0 \tev$, {\sc pythia}\cite{pythia} was used to 
simulate the effects of quark 
compositeness to leading order (LO).
The results for composite quarks relative to 
expectations from the standard model are also shown
in Fig. \ref{dth_figure} for $\Lambda_{\it LL} = 1.7, 2.0$ and $2.5 \tev$.
The ratios are independent 
of the {\sc pythia} renormalization scale for the range considered here.
Using the above ratio 
from {\sc pythia}, we scaled the {\sc jetrad} calculation 
for each PDF to obtain our estimate of the
expected cross section for any given $\Lambda_{\it LL}$.

%PAGE 9, PARA 1
As seen in Fig. \ref{dth_figure}, quark compositeness would show up 
as a relative rise in the cross section as a function of \htran.  
Changes in renormalization scale affect the absolute cross section, but 
not the shape of \htran\ distribution.
Cross sections calculated using CTEQ4M or MRST PDFs 
differ in normalization but only slightly in shape. 
Our analysis will therefore be based on comparison of the shapes of
the measured and predicted \htran\ distributions.

%PAGE 9, PARA 2
The event efficiency depends weakly on \htran, and the corrections are 
applied directly to the Monte Carlo generated events.  
The jet energies in the Monte Carlo are smeared according to measured 
resolution functions.  The effect of this smearing is also found to be
independent of \htran, resulting in just an overall rescaling of 
the \htran\ distribution.  
Finally, the jet energy scale  (and its uncertainty) is used to correct
the Monte Carlo and to determine bin-to-bin correlations in \htran.  
The expected distribution, with a variable normalization, is then compared 
directly to data.

%PAGE 9, PARA 3
The error bars in Fig. \ref{dth_figure} are statistical, and the envelope 
indicates the systematic uncertainty (one standard deviation) 
from the jet energy scale.
The systematic uncertainties range from 17\% at 
the lowest bin shown, to 34\% at the highest \htran\ bin.  Because
these uncertainties are highly correlated ($> 92$\%) 
in \htran, the line-shape of the \htran\ distribution
is quite constrained within the 95\% confidence level
(CL) limit.
%The distribution of $(Data - {\sc jetrad})/{\sc jetrad}$ in 
%Fig. \ref{dth_figure} exhibits no deviation from QCD, 
%and we see no evidence for an excess above the theoretical prediction.
%In addition, the excess in the \htran\ spectrum reported in \cite{cdf_ht} 
%is not corroborated by our measurement
%of $(Data - {\sc jetrad})/{\sc jetrad}$ in Fig. \ref{dth_figure}.
%However, the CDF and \D0 measurements are in qualitative agreement given
%the size of their respective systematic uncertainties.
%We conclude that there is 
%no evidence for quark compositeness below an energy scale
%of about $2.0 \tev$.
The distribution of $($Data$ - {\sc jetrad})/{\sc jetrad}$ in 
Fig. \ref{dth_figure} exhibits no deviation from QCD.
From this measurement, we conclude that there is 
no evidence for quark compositeness below an energy scale
of $2.0 \tev$.

%PAGE 10, PARA 1
A modified Bayesian\cite{bayesian,bayesian2} procedure 
%A modified Bayesian\cite{bayesian} procedure 
sets the 95\% CL 
lower limits on quark compositeness.  The procedure considers
the efficiencies, the smearing of jet energy in the Monte Carlo,
the integrated luminosity, 
the uncertainty and correlations on the jet energy scale, and the
normalization on the expected cross section.  Because the efficiencies,
resolutions, and integrated luminosity are independent of \htran, these
parameters were included in the normalization, which was defined to have
a flat prior probability.  A Gaussian prior was assumed 
for the jet energy scale, and a flat prior 
for $\xi \equiv 1/\Lambda_{\it LL}^2$.  The standard model 
corresponds to $\Lambda_{\it LL} \rightarrow \infty$ ($\xi \rightarrow 0$).  
The renormalization scale was varied and the results are summarized in 
Table \ref{composite_limits}.  The 95\% confidence level limits are 
obtained from the $\xi$ distributions by integrating 
the posterior probability and requiring that
95\% of the integral be below the limit.
Separate limits for both signs of the interference 
term and for the two PDFs, 
CTEQ4M and MRST, are listed in Table \ref{composite_limits}.
In general, the limits show small increases for the 
negative sign of the interference term, and the MRST PDF.
The limits also slightly increase with increasing renormalization
scale.

%PAGE 10, PARA 2
We checked the stability of the limits 
given in Table \ref{composite_limits}.  
The cut $\vert \eta^j \vert \leq 3$
was tightened to $\vert \eta^j \vert \leq 2$, thereby excluding 
events with forward jets in
the $H_T$ distribution, with essentially no impact on the limits.
Possible bias introduced by our selection of
the hard-scattering vertex was studied again, with no observed impact
on the limits.  
The \ET\ threshold of the jets was 
increased from $20 \gev$ to $50 \gev$, and 
the analysis repeated.  
The resulting limits were consistent with those 
based on the $20 \gev$ threshold.  
Changing the assumed jet energy resolution by $\pm 1$ 
standard deviation had little effect on 
the shape of the \htran\ distribution, and thus, little effect on the limit.  
Varying $\alpha_s$ was investigated through use of the CTEQ4A1-A5 
PDFs for a single choice of $\mu$ and ${\cal A}$,
as shown in Table \ref{limits-check-table}.
There is very 
little change of the limit for $0.110 \leq \alpha_s \leq 0.122$,
corresponding to a $Q^2$ range from $(50~\gev)^2$ to $(230~\gev)^2$.
The impact of the gluon content of the proton was
studied using the PDF MRST($\gup$) (one standard deviation high) 
and MRST($\gdown$) (one standard deviation low) \cite{mrst}.  The limits shown in 
Table \ref{limits-check-table} depend only weakly on this choice.  
Finally, the distribution from {\sc jetrad} (number of events in 
each \htran-bin) was 
fluctuated according to Poisson statistics, and the limit recalculated.  
The resulting limits were only $0.1 \tev$ higher than the limits 
based on the data, providing a measure of 
the sensitivity of this analysis to the finite statistics and 
uncertainties in energy-scale.

%PAGE 11, PARA 1
In summary, the measured \htran\ distribution above $500 \gev$ is well 
modeled by the {\sc jetrad} (NLO QCD) event generator. We
find no evidence for compositeness in quarks, and set lower limits 
on the compositeness scale as a function 
of renormalization scale, sign of the interference term in the 
compositeness Lagrangian, and choice of PDF.  These limits are
not affected by small variations in our analysis procedures.
The average radius of the scattered quark
(principally from the first family)
is therefore less
than $\Delta x \approx \hbar c / \Lambda_{\it LL} \approx 
1 \times 10^{-4}$ fm.

%PAGE 11, PARA 2 -- Acknowlegements para.
\vspace{.2cm}
%\input{acknowledgement_paragraph.tex}
%
% Acknowledgement_paragraph.tex
%
We thank the staffs at Fermilab and at collaborating institutions 
for contributions to this work, and acknowledge support from the 
Department of Energy and National Science Foundation (USA),  
Commissariat  \` a L'Energie Atomique and
CNRS/Institut National de Physique Nucl\'eaire et 
de Physique des Particules (France), 
Ministry for Science and Technology and Ministry for Atomic 
   Energy (Russia),
CAPES and CNPq (Brazil),
Departments of Atomic Energy and Science and Education (India),
Colciencias (Colombia),
CONACyT (Mexico),
Ministry of Education and KOSEF (Korea),
CONICET and UBACyT (Argentina),
A.P. Sloan Foundation,
and the Humboldt Foundation.

\pagebreak

%table 1
\begin{table}[ht]
\begin{center}
\caption{Previous 95\% CL limits, given in {\rm TeV}, on the left-left
isoscalar quark compositeness model.}
\label{limits_to_date}
\vspace{0.25cm}
\begin{tabular}{lcc}
Method& $\Lambda_{\it LL}^+$ & $\Lambda_{\it LL}^-$ \\ \hline
Dijet Mass (\D0) \cite{Mjj} & 2.4 & 2.7\\
Dijet Angular Distribution (\D0) \cite{d0_dijet} & 2.1 & 2.2\\
Dijet Angular Distribution (CDF) \cite{cdf_dijet} & 1.8 & 1.6\\
\end{tabular}
\end{center}
\end{table}

%table 2
\begin{table}[ht]
\begin{center}
\caption{The 95\% CL lower limits on quark compositeness in
${\rm TeV}$,
for both CTEQ4M and MRST PDFs, and for renormalization scales 
$\mu = f_E \cdot E_T^{\rm max}$ and $\mu = f_H \cdot H_T$ 
(where $E_T^{\rm max}$ is for the leading jet).  For each PDF, 
the first limit is for ${\cal A} = +1$ and 
the second is for ${\cal A} = -1$.}
\label{composite_limits}
\vspace{0.25cm}
\begin{tabular}{cccccc}
$f_E$ & CTEQ4M & MRST & $f_H$ & CTEQ4M & MRST \\
     & $\Lambda_{\it LL}^+ ~~~ \Lambda_{\it LL}^-$ 
     & $\Lambda_{\it LL}^+ ~~~\Lambda_{\it LL}^-$ & 
     & $\Lambda_{\it LL}^+ ~~~ \Lambda_{\it LL}^-$ & 
     $\Lambda_{\it LL}^+ ~~~\Lambda_{\it LL}^-$ \\ \hline
0.25&1.9 \quad 1.9&1.9 \quad 2.0&0.25&1.9 \quad 2.0&2.0 \quad 2.1 \\
0.50&1.9 \quad 2.0&2.0 \quad 2.1&0.50&2.0 \quad 2.0&2.1 \quad 2.2 \\
0.75&2.0 \quad 2.0&2.0 \quad 2.1&0.75&2.0 \quad 2.1&2.1 \quad 2.2 \\
1.00&2.0 \quad 2.0&2.1 \quad 2.2&1.00&2.0 \quad 2.1&2.1 \quad 2.2 \\
1.25&2.0 \quad 2.0&2.1 \quad 2.2&&& \\
1.50&2.0 \quad 2.1&2.1 \quad 2.2&&&
\end{tabular}
\end{center}
\end{table}

%table 3
\begin{table}[ht]
\begin{center}
\caption{The 95\% CL lower limits on quark compositeness 
scale in $\Lambda_{\it LL} ({\rm TeV})$
for different $\alpha_s$ (CTEQ4A1-5) and different gluon content 
(MRST($\gup$) and MRST($\gdown$)).  
The renormalization scale is $E_T^{\rm max}/2$, ${\cal A} = +1$.  The limits 
for CTEQ4M and MRST are included for comparison.}
\label{limits-check-table}
\vspace{0.25cm}
\begin{tabular}{c cc cc c}
PDF&$\Lambda_{\it LL}^+$&PDF&
$\Lambda_{\it LL}^+$&PDF&$\Lambda_{\it LL}^+$\\ \hline
CTEQ4A1&2.0&CTEQ4A2&2.0&CTEQ4M&1.9 \\
CTEQ4A4&1.9&CTEQ4A5&1.9&& \\
MRST($\gup$)&2.0&MRST($\gdown$)&2.1&MRST&2.0 
\end{tabular}
\vspace{-0.25cm}
\end{center}
\end{table}

%figure 1
\setlength{\unitlength}{0.7mm}
\begin{figure}[hbt]
\begin{picture}(100,150)(-40,0)
\mbox{\epsfxsize11.0cm\epsffile{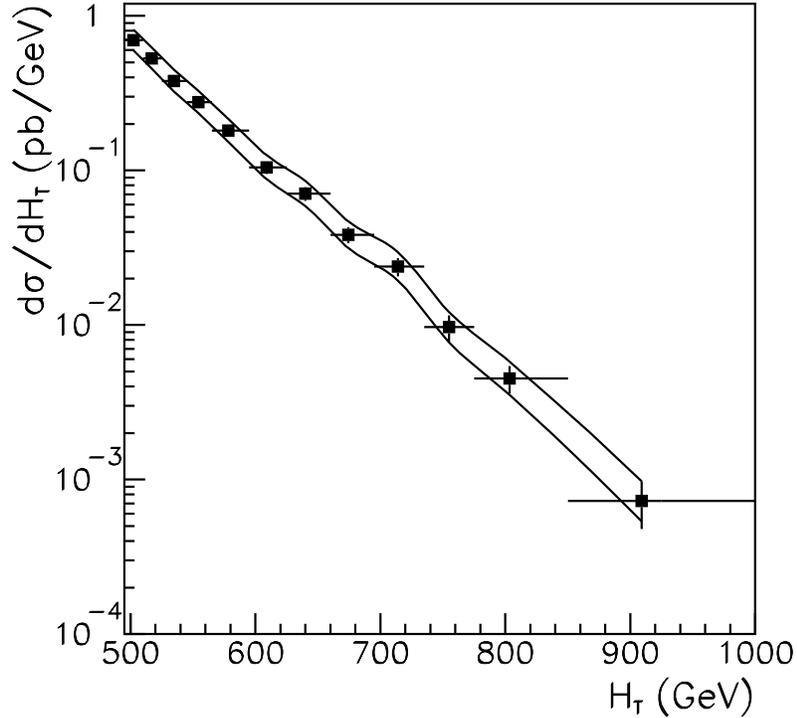}}
\end{picture}
\caption{The \htran\ distribution for \htran\ above $500 \gev$.
Error bars are statistical, and the error envelope shows the 
systematic error on the jet-energy scale.  This cross section is corrected
for efficiencies and jet energy scale, but not for resolution.}
\label{dataxsection_figure}
\end{figure}

%figure 2
\setlength{\unitlength}{0.7mm}
\begin{figure}[hbt]
\begin{picture}(100,150)(-40,0)
\mbox{\epsfxsize11.0cm\epsffile{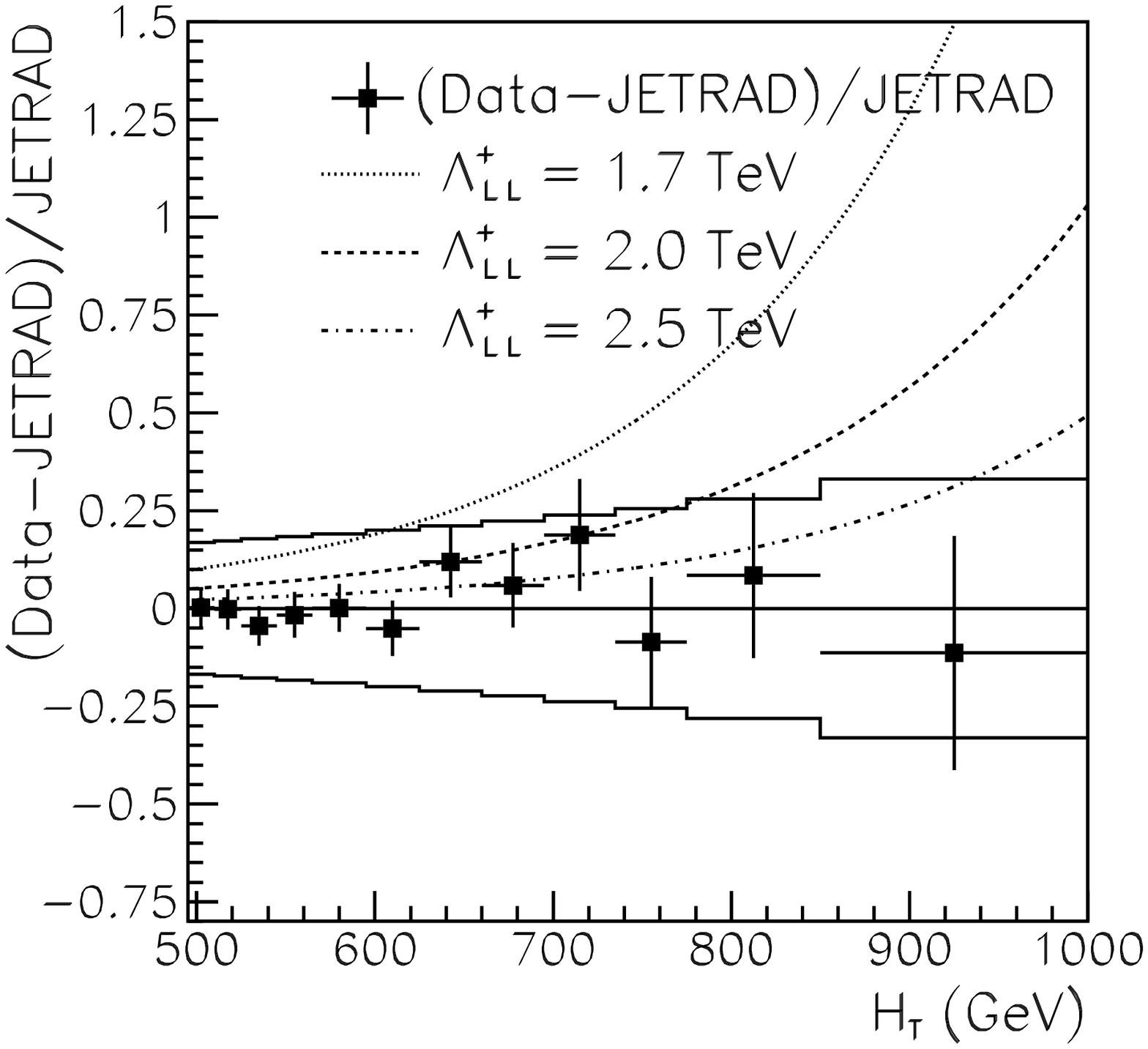}}
\end{picture}
\caption{Comparison of the measured \htran\ distribution with 
{\sc jetrad} (CTEQ4M and a
renormalization scale of $\mu = E_T^{\rm max}/2$).  
The errors on the points are statistical, and the 
error band represents the highly correlated 
systematic uncertainty due to the jet energy 
scale.  The superimposed 
curves correspond to expectations for three compositeness scales.}
\label{dth_figure}
\end{figure}


\begin{references}

\bibitem{Rutherford}
E. Rutherford, J. Chadwick, C. D. Ellis (1930). 
{\it Radiations from radioactive substances.}
Cambridge University Press.

\bibitem{Baden}
D. Baden,
Int.J.Mod.Phys. {\bf A13}, 1817 (1998).

\bibitem{d0_inc}
\D0\ Collaboration, B. Abbott, {\it et al.},
Phys. Rev. Lett. {\bf 82}, 2451 (1999),
hep-ex/9807018.
%Fermilab-Pub-98/207-E.

\bibitem{cdf_inc}
\CDF\ Collaboration, F. Abe {\it et al.,}
Phys. Rev. Lett. {\bf 77}, 438 (1996),
hep-ex/9601008.

\bibitem{Mjj}
\D0\ Collaboration, B. Abbott {\it et al.,}
Phys. Rev. Lett. {\bf 82}, 2457 (1999), 
hep-ex/9807014.

\bibitem{d0_dijet}
\D0\ Collaboration, B. Abbott {\it et al.,}
Phys. Rev. Lett. {\bf 80}, 666 (1998),
hep-ex/9707016.

\bibitem{cdf_dijet}
\CDF\ Collaboration, F. Abe {\it et al.},
Phys. Rev. Lett. {\bf 77}, 5336 (1996).  Erratum - {\em ibid.}
 {\bf 78}, 4307 (1997),
hep-ex/9609011.

\bibitem{cdf_ht}
\CDF\ Collaboration, F. Abe {\it et al.},
Phys. Rev. Lett. {\bf 80}, 3461 (1998),
Fermilab-Pub-97-290-E.

\bibitem{theory-comp}
E. Eichten, K. Lane and M.E. Peskin,
Phys. Rev. Letts. {\bf 50}, 811 (1983); 
E. Eichten, I. Hinchliffe, K. Lane, and C. Quigg,
Rev. Mod. Phys. {\bf 56}, 579 (1984);
ibid., {\bf 58}, 1065 (1986);
K. Lane, hep-ph/9605257.

\bibitem{det}
\D0\ Collaboration, S. Abachi {\it et al.},
Nucl. Instr. \& Meth. {\bf A338}, 185 (1994),
Fermilab-Pub-93-179-E.

\bibitem{luminosity}
J. Bantly et al., Fermilab-TM-1995 (unpublished). In order to
facilitate comparison with previously published results, this analysis does
not use the luminosity normalization given in the Dzero Collaboration, B.
Abbott et al., Phys. Rev. {\bf D61}, 072001 (2000), sec. VII, 
hep-ex/9906025. The updated normalization would have the effect 
of increasing the luminosity by 3.2\%.

\bibitem{d0_escale}
\D0\ Collaboration, B. Abbott {\it et al.,} 
Nucl. Instr. \& Meth. {\bf A424}, 352 (1999),
hep-ex/9805009.

\bibitem{jetrad}
W. Giele, E. Glover, and D. Kosower,
Nucl. Phys. {\bf B403}, 633 (1993),
hep-ph/9302225.

\bibitem{cteq}
H.L. Lai {\it et al.,}
Phys. Rev. D {\bf 55}, 1280 (1997),
hep-ph/9606399.

\bibitem{mrst}
A.D. Martin {\it  et al.,} 
Eur.\ Phys.\ J. C {\bf 4} 463 (1998), 
hep-ph/9803445.

\bibitem{pythia}
T. Sj\"ostrand, 
Comp. Phys. Comm. {\bf 82}, 74 (1994).
{\sc pythia 5.7} only contains
the left-left isoscalar model of quark compositeness.

% \bibitem{theory-comp-corr}
% K. Lane,
% BUHEP-96-8, hep-ph/9605257 (1996) and private communication.
               
\bibitem{bayesian}
F.T. Solmitz,
Ann. Rev. Nucl. Sci.  {\bf 14}, 375 (1964).

\bibitem{bayesian2}
I. Bertram, {\it et al.,} Fermilab-TM-2104 (unpublished).

\end{references}
\end{document}